# Efficient UAV Physical Layer Security based on Deep Learning and Artificial Noise

Behrooz Khadem, Department of Communication,
Imam Hossein Comprehensive University, Tehran, Iran
Salar Mohebalizadeh, Department of Computer and Electrical
Engineering, University of Tabriz, Tabriz, Iran

*Abstract*—Network-connected unmanned aerial vehicle (UAV) communications is a common solution to achieve high-rate image transmission. The broadcast nature of these wireless networks makes this communication vulnerable to eavesdropping. This paper considers the problem of compressed secret image transmission between two nodes, in the presence of a passive eavesdropper. In this paper, we use auto encoder/decoder convolutional neural networks, which by using deep learning algorithms, allow us to compress/un-compress images. Also we use network physical layer features to generate high rate artificial noise to secure the data. Using features of the channel with applying artificial noises, reduce the channel capacity of the unauthorized users and prevent eavesdropper from detecting received data. Our simulation experiments show that for received data with SNR fewer than 5 in the authorized node, the MSE is less than 0.05.

## I. INTRODUCTION

UAVs with their high mobility and low cost, also commonly known as drones, have found a wide range of applications during the past few decades. Historically, UAVs have been primarily used in the military, mainly deployed in hostile territory to reduce pilot losses. With continuous cost reduction and device miniaturization, small UAVs are now more easily accessible to the public; hence, numerous new applications in the civilian and commercial domains have emerged, with typical examples including weather monitoring, forest fire detection, traffic control, cargo transport, emergency search and rescue, communication relaying, and others [1, 2].

Nowadays, neural networks and machine learning especially deep learning are used for widespread aims such as image processing, signal processing and transmitting data in UAV networks [3,4,5,6]. In this paper, an efficient secure FANET [7] approach for compressed image transmission is presented. Since classical cryptography has several weaknesses against ordinary and side channel attacks [8], physical layer gives an ability to propose secure solutions for data transmission [9]. Figure (1) shows a sample UAV Ad-hoc network. In these networks, there is a backbone UAV, which other nodes send their data to it, then it sends data to the ground station[10].

## II. RELATED WORKS

Several works have proposed some techniques to increase the security and efficiency of UAV data transmission.
Masci et al. proposed an unsupervised learning method that is an auto CE in order to use image compression and data coding [11].
One of the challenges of using JPEG2000 was solved by Wang and colleagues, they used deep learning for solving the problem and tried to improve the quality of the compressed image [12].

Geoger et al. used reversible and developed deep learning algorithms to generate compressed data. They collected widespread data and used abstract features; hence they showed Unsupervised neural networks are better for UAVs because UAVs have special limits such as energy and time saving as well as UAVs should do their duty independently, therefore, unsupervised learning algorithms are better for UAVs because these algorithms don't need labels like supervised algorithms [13].

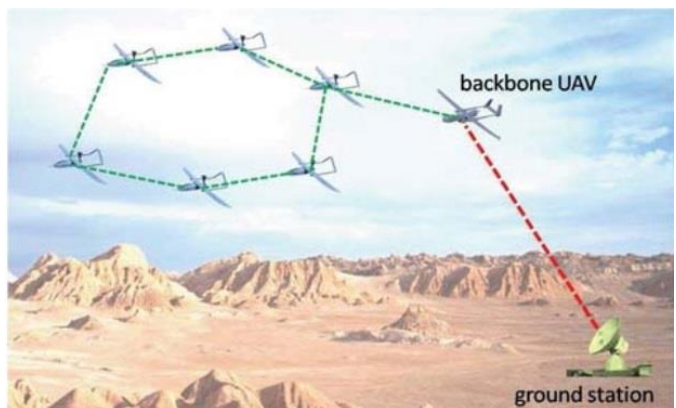

Figure 1: UAV Ad-hoc network [12]

Extracted features in the deep learning algorithms are used for transmission of the collected data, especially in UAV networks. The deep learning algorithm can obtain high-value features automatically and we do not need prior methods for extracting features. The features that are extracted in deep learning algorithms are more efficient than previous features. When deep learning is used for the recovering images, extracted data is similar to general images; this issue is one of the benefits of the deep learning. The Deep learning algorithm eliminates weak features through training progress, this operation makes the algorithm not to select weak features, therefore, better features are selected and images are recovered accurately [13].

In this paper, we have used both an auto encoder/decoder deep learning algorithm that has utilized a CNN [14]. On the other hand, we have gained from deconvolutional networks for decoding [15]. The auto encoder and decoder methods are trained by a database that is called training data, this approach is the unsupervised method for training data. Encoder step receives data and transfers them to latent codes then latent codes are sent to the decoder and it received latent codes. This step recovers primary data from latent codes. The decoder does this duty by using de-convolutional networks.

Researchers in aforementioned works ignored the fact that there may be an Eav in the network which it receives data and uses the gain of the network. In such a case, when the deep learning is used to prepare data , if Eav knows the training algorithm and other information about learning method, she would be able to detect data and decode them [13].Traditionally, this attack is one of the passive attacks defined in the Ad-hoc networks. Commonly, cryptography algorithms are used in order to secure wireless communication networks. But these algorithms have some challenges as the follows [16]:

- Computational limits in wireless nodes [17].
- Bounded battery power in nodes [17].
- Difficulty of the distributed nodes management in the form of centralized [17].
- Key producing and management problems exist for big networks [18, 19]

Recently some solutions are presented for preventing from eavesdropping attack:

- Designing high-energy consuming cryptography algorithms for encrypting communication [20-22]
- Using power control and directional antenna to reduce eavesdropping possibility [23, 24]

Traditionally, secure key based cryptography techniques are used in order to prevent from eavesdropping [25]. Key generation and management need a reliable infrastructure, which has some challenges in UAV networks and Ad-hoc networks [26]. An alternative method suggests that the physical layer features can help us in order to provide communication security. This method is a novel and Influential one which is used for preventing from Eav attacks, in both wireless network and wiretap channels [27]. Physical layer security is a keyless solution which can be used in decentralized massive networks [28].

Shannon and Siezar showed that because of difference between the authorized channel and unauthorized one, if they could reduce the Eav channel capacity, they would send data securely [29, 30]. If the Eav's unauthorized-channel capacity is bigger than the authorized-channel capacity, the security capacity will be zero [29].

The main contribution of this paper is to present an efficient secure solution for real-time image transmission in UAV networks. Our solution consists of two main objectives. First we use artificial noise in order to reduce the Eav's channel capacity such that added artificial noises have not influenced the authorized channel capacity (since these noises produced such that lies in the null space of the authorized channel). Second we use deep learning so far to compress and reduce the data for real-time applications in UAV networks.

Section 2 introduces basic concepts. Section 3 provides our proposed solution for efficient secured compressed image transmission for UAV networks, section 4 shows the results of simulations and section 5 contains our conclusion and future recommendations.

### III. BASIC CONCEPTS

The basic concept of this paper is deep neural network which is introduced briefly in the following. Table (1) shows some used abbreviations in this paper.

TABLE 1: ABBREVIASIONS

| | |
|---|---|
| Auto-Encoder | AE |
| Bit Error Rate | BER |
| Convolutional Encoder | CE |
| Convolutional Auto-Encoder | CAE |
| Convolutional Neural Network | CNN |
| Deconvolutional Decoder | DD |
| Deconvolutional Network | DN |
| Eavesdropper | Eav |
| Flying Ad-hoc Network | FANET |
| Mobile Ad-hoc Network | MANET |
| Mean squared normalized error | MSE |
| Peak Signal to Noise Ratio | PSNR |
| Signal to Noise Ratio | SNR |

Neural network often includes several steps as data collection, network construction, network configuration, weight vectors definition and bias setup, network training, and finally network implementation [31]. Generally, neural networks are divided into two main categories, one layer and multi-layer networks. One layer network has a weight layer and provides two groups of neural cells, input neural cells, and output neural cells. Input neural cells receive data and output neural cells generate results. The multi-layer network has some layers between the input and output layers which are called hidden layers [32].

Lately, deep learning becomes a rapidly expanding research topic, which is part of a broader family of machine learning methods based on learning representations [31, 39]. Deep learning can be generally understood as deep neural networks with multiple nonlinear layers, in which the features are learned from data through a general-purpose learning procedure, but not designed by human engineers [38]. Deep learning can solve both linear and nonlinear problems. One of the most famous breakthroughs made by deep learning was a computer program AlphaGo from Google DeepMind, who beat a professional player at the board game Go for the first time [36]. Additionally, as a current research hotspot, deep learning is making major advances in a wide variety of applications, such as image recognition based on CNN [37]. In this paper, we use feature extraction for real-time compressing images and it is based on taking advantages of deep learning. Also in this paper we utilize artificial noises that are introduced by Negi and Goel, which is a powerful approach that gives us the ability to communicate secure data among nodes without keys [33].

## IV. PROPOSED METHOD FOR SECURE-REAL-TIME DATA TRANSMISSION

Our proposed idea provides three parts. Part A provides the way we use for preparing and compressing data. In this part we illustrate the method which is used for extracting features and compressing data, using AE. Part B illustrates how we add artificial noises to encoded data so that it cannot influence on authored channel capacity. Our final combined solution using part A and part B is presented in part C to prevent Eav from eavesdropping.

### A. A framework for data preparation based on deep learning

One of the challenges in UAV networks is to transmit secure real-time data as well as pprotecting the quality of the recovered data. In one of the prior research, the real-time data transmission is presented but they ignore the security of the transmitted data.

Figure (2) shows the structure of end to end, encoding and decoding data transmission. Also Figure (2) shows that how the encoder step (CE) maps input data to latent codes with constant length. Collected data are coded in convolutional step and transmitted to latent codes. The length of latent codes is selected by the user who should make a balance between energy consumption and quality of the recovered data. The eexperiments showed that the better the retrieved data are obtained if the longer code length and more power consumption are used. As Figure (2) shows CE step has three layers, two convolutional layers, and a dense layer for connecting different layers and neural cells together [5].

Any of the convolutional layers has 16 filters with $5 \times 5$ length; these filters are used for creating 16 features. Extracted features in the first layer are entranced to the second similar layer. The last layer is a dense layer that generates latent codes which their length is selected before training [31]. output data is generated based on dense layer dimensions [13, 31]. Operations of the first and second layer are similar to extract better features [14].

The inverse de-convolutional decoder uses DN and decodes data [5]. DD uses three convolutional layers in reverse ordering of CE. The first layer is a dense layer that uses input latent code in order to generate received feature space which has $w \times h \times 16$ dimensions. Two-second layers are inverse convolutional layers. These layers map output of the dense layer to space with $w \times h \times 3$ dimensions, to recover images from received latent codes. The goal of this method is decreasing the difference between input data and recovered data and the experiments shows that the decoding error is less than prior methods. One of the advantages of this method is to send real-time-secure data with low energy consumption. Input data are normalized and have Gaussian standard distribution [13].

### B. Secure data transmit

In FANET networks, key management and distribution have a lot of challenges, so we need a method that allows us to secure network without using symmetric or asymmetric encryption keys. An alternative method is to use the physical layer security with generating artificial noise [7, 34]. Figure (3) shows transmitter A with $N_T$ antennas, receiver B with $N_R$ antennas and Eav with $N_E$ antennas. Similarly it shows a UAV network with $N_R$ UAVs that everyone has an antenna and an Eav E with $N_E$ antennas or a group of the Eav's that everyone has an antenna.

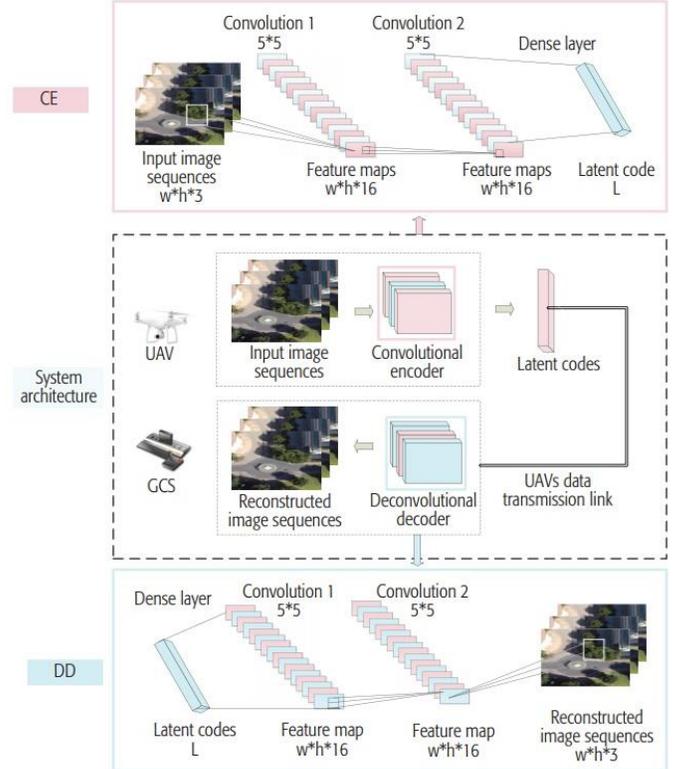

Figure 2: Structure of the encoding and decoding[13]

In worst case analysis, there are multi Eav's and everyone receives data with different channel situation. The transmitter A sends $x_k$ at time $k$, the receiver B and the Eav receive the message based on equations 1 and 2,

$$z_k = H_k x_k + n_k \qquad (1)$$
$$y_k = G_k x_k + e_k \qquad (2)$$

Where the components $n_k$ and $e_k$ are i.i.d. additive White Gaussian Noise samples with variance $\sigma_n^2$ and $\sigma_e^2$ respectively. Two matrixes, $H_k$ and $G_k$ denote the channels of the receiver and Eav at time k and are both constant and independent. Block fading is assumed and all elements of channels ($h_{i,j}$ and $g_{i,j}$) are the channel gains from transmit antenna $i$ to receive Eav antenna $j$ and are independent complex numbers. We propose that the receiver can estimate $H_k$ and sends it to the transmitter so Eav can also receive $H_k$ but not $G_k$ (Eav is passive so the amount of $G_k$ is unknown). obviously security capacity is independent of channel capacity [35]. We assume that both the receiver and the Eav have a single antenna and multiple Eav's cannot collude (i.e., $N_E = N_R = 1$). Now the concept of the artificial noise can be used. It is assumed that the receiver is able to estimate its channel $H_k$ perfectly and feed it back to the transmitter noiselessly. We assume that $H_k$ is communicated to the transmitter by an authenticated broadcast (which may be heard by the Eav). Thus, it is assumed that the Eav may know both the receiver's and its own channel. A passive Eav is assumed, which means that she only listens but does not transmit. Hence, her channel $G_k$ may not be known to the transmitter. The artificial noise is produced such that it lies in the null space of the receiver's channel, while the information signal is transmitted in the range space of the receiver's channel. This design relies on knowledge of the receiver's channel, but not of the Eav's channel. The receiver's channel nulls out the artificial noise, and hence, the receiver is not affected by the noise. However, in general, the Eav's channel will be degraded, since its range space will be different from that of the receiver's channel, and hence, some component of artificial noise will lie in its range space.

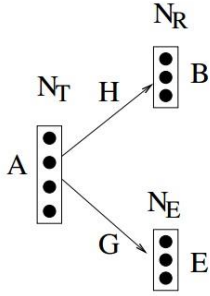

Figure 3: communication structure in the presence of Eav's[34]

The transmitter can generate artificial Noise to degrade the Eav's channel [34]. The transmitter Chooses $x_k$ as the sum of information bearing signal $s_k$ and the artificial noise signal (Equation 3).

$$x_k = s_k + w_k \quad (3)$$

Table 2: Situation of channels

|  | Channel | Coefficients | Noise | Variance of noise |
|---|---|---|---|---|
| Receiver | $H_k$ | $h_{i,j}$ | $n_k$ | $\sigma_n^2$ |
| Eav | $G_k$ | $g_{i,j}$ | $e_k$ | $\sigma_e^2$ |

Both $s_k$ and $w_k$ are assumed complex Gaussian vectors. $w_k$ is chosen to lie in the null space of $H_k$, such that $H_k . w_k = 0$. If $Z_k$ is an orthogonal basis for the null space of $H_k$, then $w_k = Z_k . V_k$ and $Z_k^\dagger Z_k = I$, therefore the signals received by the receiver and the Eav are given by, equations 4 and 5 respectively [34].

$$z_k = H_k s_k + n_k \quad (4)$$
$$y_k = G_k s_k + G_k w_k + e_k \quad (5)$$

Note how the artificial noise $w_k$ is nulled out by the receiver's Channel but not necessarily by the Eav's channel. Thus, the Eav's channel is degraded with high probability, while that of the receiver remains unaffected. If $w_k$ was chosen fixed, the artificial noise seen by the Eav would be small if $\|G_k w_k\|$ is small. To avoid this possibility, the sequence of $w_k$ is chosen to be complex Gaussian random vectors in the null space of $H_k$ [34]. In particular, the transmitter chooses elements of $v_k$ to be i.i.d. complex Gaussian random variables with variance $\sigma_v^2$.

### C. Proposed method

One of the problems in reference [13] is that we cannot send secure data. In this paper, we survey proposed methods that is presented for secure data transmit in UAV networks such as key-based approaches and keyless providing a secure space for communication, for example, physical layer. We propose using added artificial noise in order to generate a secure communication in UAV networks.

Similarity between FANET and MANET let us to use proposed method by Goel [34] in UAV networks. In other words, we can use artificial noise in order to send secure data in UAV network. When we use artificial noise and deep learning, we can send not only real-time but also secure data. In this method, secure capacity compute based on equations 6 and 7 [34].

Since $H_k$ is a vector channel, the transmitter chooses the information bearing signal as $s_k = p_k u_k$ where $u_k$ is the information signal. We assume that Gaussian codes are used. $p_k$ is chosen such that $H_k u_k \neq 0$ and $\|p_k\| = 1$. Now, secrecy capacity is bounded below by the difference in mutual information between the transmitter and the receiver versus the transmitter and the Eav,

$$Secrecy\ capacity \geq C_{sec}^a = I(Z; U) - I(Y; U) \quad (6)$$
$$= \log\left(1 + \frac{|H_k p_k|^2 \sigma_u^2}{\sigma_u^2}\right) - \log\left(1 + \frac{|G_k p_k|^2 \sigma_u^2}{E|G_k w_k|^2 + \sigma_e^2}\right) \quad (7)$$

Where $E|G_k w_k|^2 = (G_k Z_k G_k^\dagger Z_k^\dagger)\sigma_v^2$. For a passive Eav, $G_k$ is not known to the transmitter, so using the concavity of $\log(.)$ and the i.i.d. assumption of $H_k$ the average secrecy capacity is maximized by choosing $p_k = H_k^\dagger / \|H_k\|$. Thus, the information bearing signal $s_k$ lies in the range space of $H_k^\dagger$ whereas the artificial noise lies in the null space of $H_k^\dagger$. Therefore based on the aforementioned assumptions and equations and results of [34] the secrecy capacity always will be remain positive.

## V. SIMULATION RESULTS

In this paper, we try to use features of the physical layer and differences between receiver's and Eav's channels to produce and add artificial noise in the null space of the receiver channel in order to reduce channel capacity of Eav without reducing channel capacity of receiver.

In our experiments, we use digit_Train_Cell_Array_Data that is a database in Matlab for training the neural network witch its database contains 5003 images. Figure (4) shows some samples of this database.

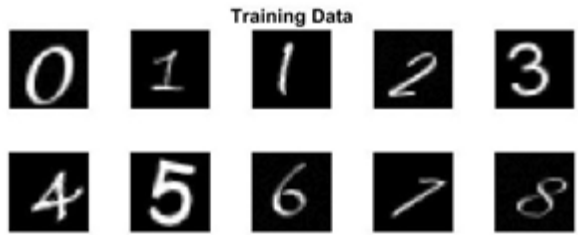
Figure 4: samples of training data.

We use training data in order to train the deep learning algorithm, after training we use test data from digit_Test_Cell_Array_Data in Matlab. This database contains 4997 images. Figure (5) shows some samples of them.

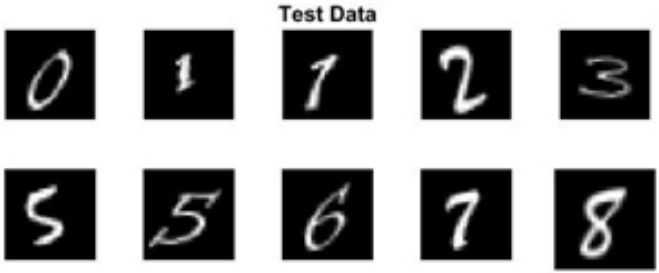
Figure 5: samples of test data

We use MSE in order to compute received data error. This function computes different between predicted data with original data based on equation 8.

$$MSE = \frac{1}{N}\sum_i \frac{(P_i - M_i)^2}{\bar{P}\bar{M}}$$
$$\bar{M} = \frac{1}{N}\sum_i M_i$$
$$\bar{P} = \frac{1}{N}\sum_i P_i \qquad (8)$$

We send encoded data with 32, 64, 128, 256 latent code layers. The number of layers shows that we use more vital data for compression data. In all situations, the number of iteration is 400 and our system is run on a system with Intel core i7-4710HQ CPU2.50 GHz and RAM 12.0GB for training deep learning algorithm. Figures (6) through (9) show received data by the receiver when latent codes layers are 32, 64, 128 and 256 respectively. Figures (10) through (13) show received data by Eav.

In these figures, we can see that if the Eav knows our deep learning algorithm, she cannot decode original data from received latent codes because there is added artificial noise on the transmitted data.

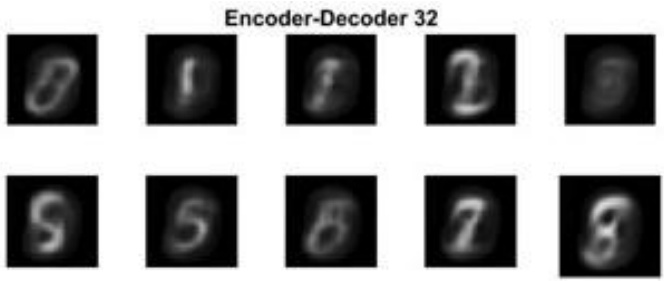
Figure 6: received data when receiver and transmitter use dense layer with 32 latent code, MSE=0.022

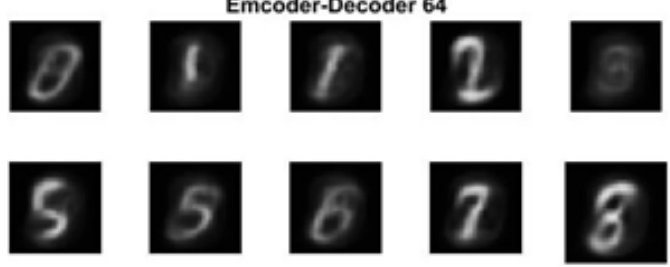
Figure 7: received data when receiver and transmitter use dense layer with 64 latent code, MSE=0.0176.

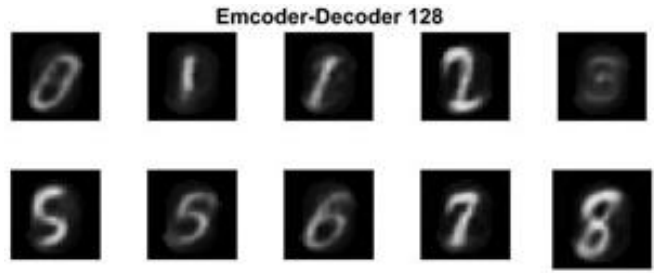
Figure 8: received data when receiver and transmitter use dense layer with 128 latent code, MSE=0.0149

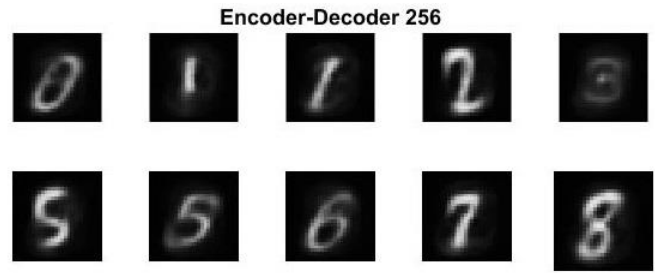
Figure 9: received data when receiver and transmitter use dense layer with 256 latent code, MSE=0.0122

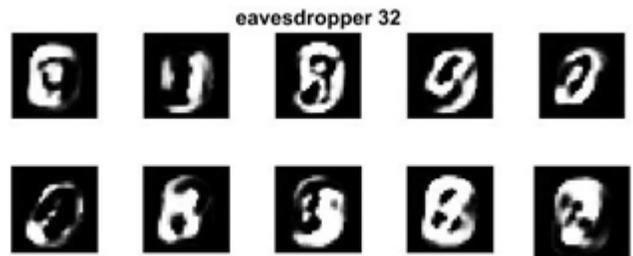
Figure 10: received data when Eav and transmitter use a dense layer with 32 latent code, MSE=0.1209.

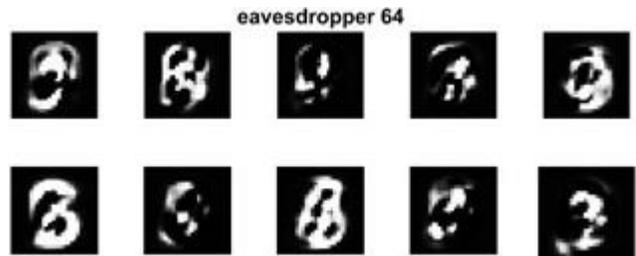
Figure 11: received data when Eav and transmitter use a dense layer with 64 latent code, MSE=0.1210.

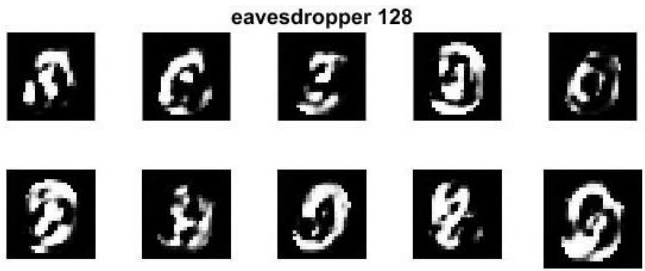

Figure 12: received data when Eav and transmitter use a dense layer with 128 latent code, MSE=0.1161.

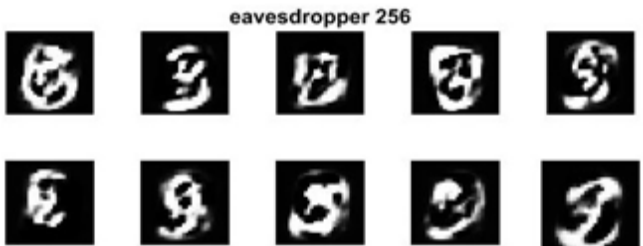

Figure 13: received data when Eav and transmitter use a dense layer with 256 latent code, MSE=0.1146.

Compression ratio and average PSNR of the decoded data are shown in Table (3).

Table 3: Compression ratio and PSNR.

| Latent code | PSNR | Compression ratio |
|---|---|---|
| 32 | 16.75 | ٪4.08 |
| 64 | 17.74 | ٪8.16 |
| 128 | 18.53 | ٪16.33 |
| 256 | 19.35 | ٪32.65 |

Figure (14) shows operation of the deep learning algorithm when we use different latent codes. Based on simulation result, it can be seen that deep learning algorithm that uses more latent codes layer has better operation than other algorithms.

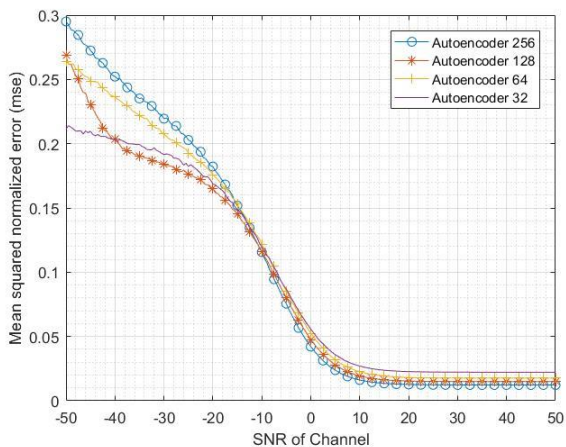

Figure 14: receives data error by receiver with different latent codes

Figures (15) through (18) compare received data error by Eav with the receiver when they use deep learning algorithms with different latent code layers. These figures show Eav can't detect data because artificial noises increase received data error.

We propose the worst situation in which Eav knows the deep learning algorithm that we use for compression data and creating latent codes. However, Eav cannot detect data because we secure them by using artificial noise.

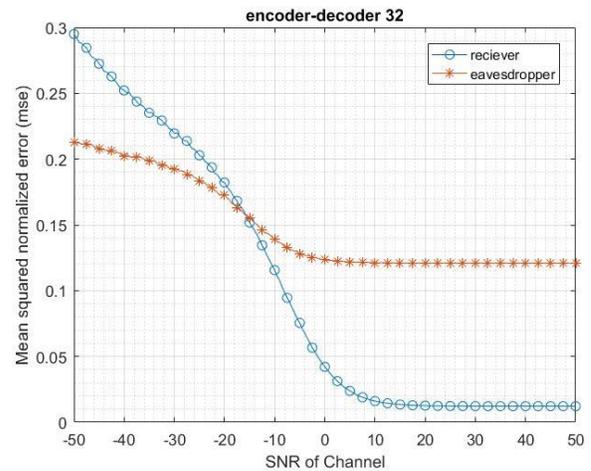

Figure 15: MSE of the received data by Eav and receiver when they use deep learning algorithms with 32 latent codes

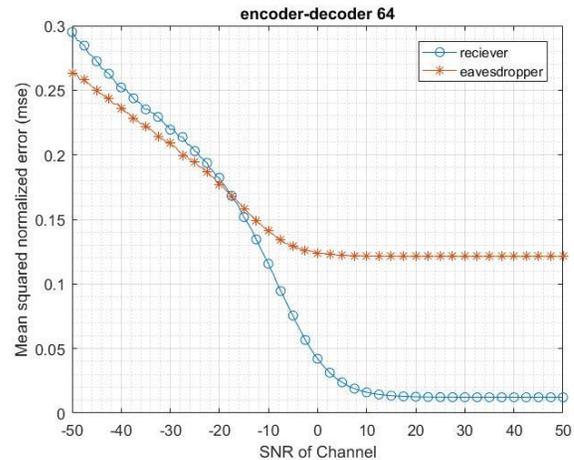

Figure 16: MSE of the received data by Eav and receiver when they use deep learning algorithms with 64 latent codes

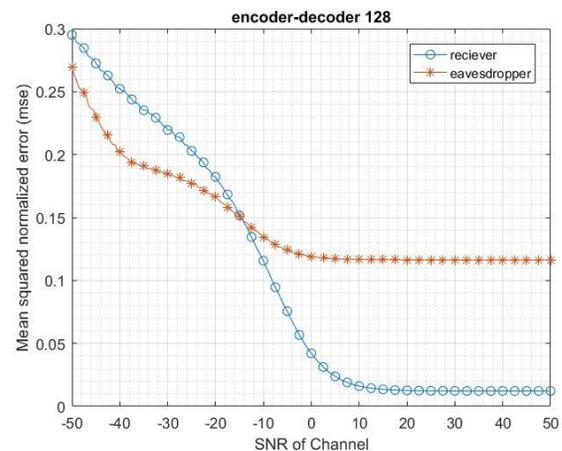

Figure 17: MSE of the received data in Eav and receiver when they use deep learning algorithms with 128 latent codes

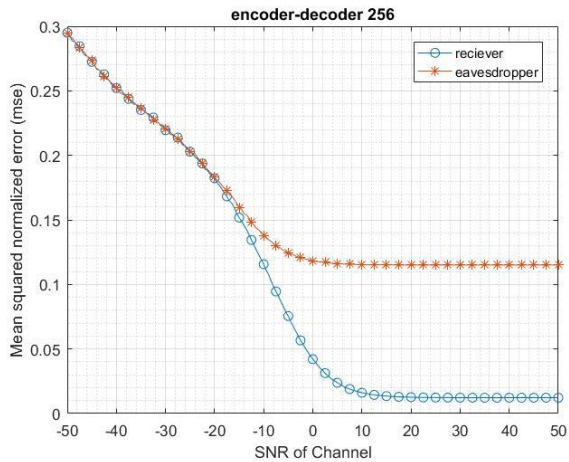

Figure 18: MSE of the received data by Eav and receiver when they use deep learning algorithms with 256 latent codes

Figure (18) shows BER in the authorized node and Eav, in order to illustrate the effect of artificial noise when data is detected. We use BPSK modulation in all cases. The channel is distributed Riley channel and there are added white Gaussian noise on data.

## VI. CONCLUSION AND FUTURE WORK

UAV networks are decentralized, therefore we cannot use key based cryptography algorithms because there are a lot of challenges. Hence, in this paper, we use physical layer feature in order to generate a secure real-time communication method that can be used in UAV networks. In this paper, we propose that there is passive Eav and for future research it can be assumed that the Eav is active.

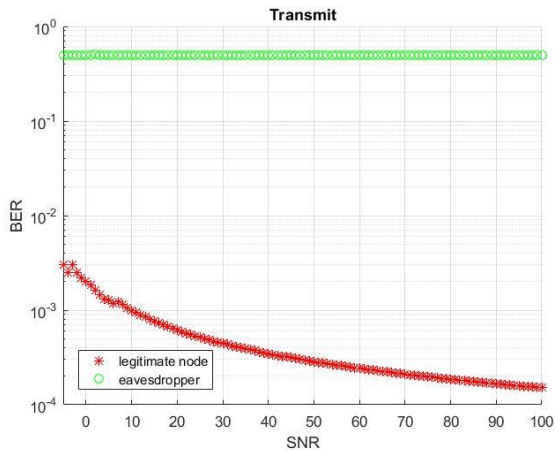

Figue19: comparing received data error between Eav and authorized node

One way to prevent from these kinds of Eav's is to estimate channel of the Eav and generate artificial noise in her channel. Based on the results of Figures (14) through (17) and Table (2) our proposed method has some advantages for instance compression ratio is selected by the user and Eav can't detect received data because they have high error amount. In this paper, we assume that the transmitter knows channel state information of the receiver, so we recommend for future work, one can try to send secure data without knowing channel state information of the receiver. Also one can use better learning algorithms for compression images and extracting features.